\documentclass[11pt]{article}
\usepackage{moriond2000,epsfig}

\def\MNRAS{\textit{ Mon.\ Not.\ Roy.\ Astr.\ Soc.\ }}
\def\ApJ{\textit{ Astroph.\ Journ.\ }}

\def\AA{\textit{ Astron.\ Astroph.\ }}

\def\be{\begin{equation}}
\def\ee{\end{equation}}
\def\bea{\begin{eqnarray}}
\def\eea{\end{eqnarray}}

\def\la{\mathrel{\mathchoice {\vcenter{\offinterlineskip\halign{\hfil
$\displaystyle##$\hfil\cr<\cr\sim\cr}}}
{\vcenter{\offinterlineskip\halign{\hfil$\textstyle##$\hfil\cr
<\cr\sim\cr}}}
{\vcenter{\offinterlineskip\halign{\hfil$\scriptstyle##$\hfil\cr
<\cr\sim\cr}}}
{\vcenter{\offinterlineskip\halign{\hfil$\scriptscriptstyle##$\hfil\cr
<\cr\sim\cr}}}}}

\def\ga{\mathrel{\mathchoice {\vcenter{\offinterlineskip\halign{\hfil
$\displaystyle##$\hfil\cr>\cr\sim\cr}}}
{\vcenter{\offinterlineskip\halign{\hfil$\textstyle##$\hfil\cr
>\cr\sim\cr}}}
{\vcenter{\offinterlineskip\halign{\hfil$\scriptstyle##$\hfil\cr
>\cr\sim\cr}}}
{\vcenter{\offinterlineskip\halign{\hfil$\scriptscriptstyle##$\hfil\cr
>\cr\sim\cr}}}}}

\begin{document}
\vspace*{4cm}
\title{A combined HST/CFH12k/XMM survey of X-ray luminous clusters of
  galaxies at \boldmath{$z\sim0.2$}}

\author{OLIVER CZOSKE$^1$, JEAN--PAUL KNEIB$^1$, IAN SMAIL$^2$, GRAHAM
  P. SMITH$^2$, HARALD EBELING$^3$}

\address{$^1$Laboratoire d'Astrophysique, Observatoire Midi--Pyr\'en\'ees,
  14 av.\ Ed.\ Belin, 31400 Toulouse, France\\
  $^2$Department of Physics, University of Durham, South Road, Durham
  DH1 3LE, UK\\
  $^3$Institute for Astronomy, University of Hawai`i, 2680 Woodlawn
  Drive, Honolulu, HI96822, USA}

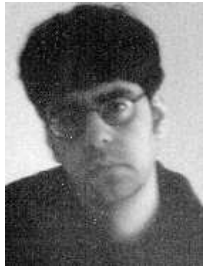
\maketitle\abstracts{
  We describe a project to study a sample of X--ray luminous clusters
  of galaxies at redshift $z\sim0.2$ at several scales (with HST/WFPC2 and
  CFHT/CFH12k) and wavebands (optical and X--ray). The main aims of the
  project are (i) to determine the mass 
  profiles of the clusters on scales ranging from $\sim
  10\,h^{-1}\,\mbox{kpc}$ to $\ga1.5\,h^{-1}\,\mbox{Mpc}$ using weak
  and strong lensing, thereby testing theoretical predictions of a
  ``universal mass profile'', and (ii) to calibrate the $M_{\rm
  total}$---$T_{\rm X}$ relation in view of future application in the
  study of the evolution of the cluster mass function at higher redshift. 
  }

\section{Introduction}
Current models describe the formation and evolution of large--scale
structure in the Universe in a hierarchical, ``bottom--up'',
way. Since clusters of galaxies are the most massive gravitationally
bound objects found in the Universe at the present time, their
evolution is observable at low redshift, $z\!\la\! 1$. Clusters of
galaxies are therefore powerful probes for testing cosmological
scenarios and determining cosmological parameters. 

Two theoretical predictions are of particular interest. The first one
concerns the evolution of the cluster mass function with redshift,
which can be described using linear theory. As shown by Eke et
al.~\cite{eke}, in a low matter density universe ($\Omega_{\rm M} \sim
0.3$) we expect to see about 30 times as many clusters out
to $z=1$ as in a high matter density universe ($\Omega_{\rm M} =
1$), using the current cluster abundance to normalize the density
fluctuation power spectrum.  

A second theoretical prediction concerns the internal structure of
clusters. Numerical simulations suggest that dark matter haloes over a
wide range of masses can be accurately described by a universal mass
profile~\cite{nfw}. In the context of numerical simulations this is a
very robust prediction, as the profile works over a wide range of
masses, radii and cosmological parameters. Other groups, however, find
different behaviour of the mass distribution in their simulations,
notably in the centres of haloes~\cite{moore}. 

The obserational tool of choice for investigating cluster mass
profiles is gravitational lensing, which in principle permits to
access the mass distribution directly (albeit a weighted sum of all
the mass between the observer and the source plane). 

Even given the large upcoming CCD mosaic cameras (such as MEGACAM),
large--area cluster searches (necessary for testing predictions
concerning the evolution of the cluster mass functio) using their weak
lensing signature will be difficult at best. More practical will be
cluster searches from X--ray all--sky surveys. The temperature $T_{\rm
  X}$ of the hot intracluster gas is directly related to the depth of
the gravitational potential of the cluster if the gas is in
hydrostatic equilibrium. In that case the cluster mass function can be
transformed into a temperature function which retains the large
separation between high-- and low--$\Omega_{\rm M}$
universes~\cite{eke}.  However, current observations indicate that the
$M_{\rm total}$---$T_{\rm X}$ relation differs from simple theoretical
predictions, possibly related to pre--heating of the
gas~\cite{bower}. Also the impact of substructure in the mass and
gas distribution within clusters on the scatter around the $M_{\rm
  total}$---$T_{\rm X}$ relation has yet to be studied in a systematic
way. It is therefore imperative to calibrate the shape of and the
scatter around the $M_{\rm total}$---$T_{\rm X}$ relation
observationally on a well--defined sample of clusters of galaxies
before using it to test cosmological predictions. 

Our project aims at studying a sample of massive galaxies at redshift $z \sim
0.2$ using HST and CFH12k observations in order to constrain mass
profiles on length scales between $\sim 10 h^{-1}$~kpc out to $\ga1.5
h^{-1}$~Mpc and to relate the lensing masses to X--ray observables
(notably $T_{\rm X}$) from observations with XMM--Newton. 

\section{Sample selection}

\begin{figure}
  \begin{center}
    \psfig{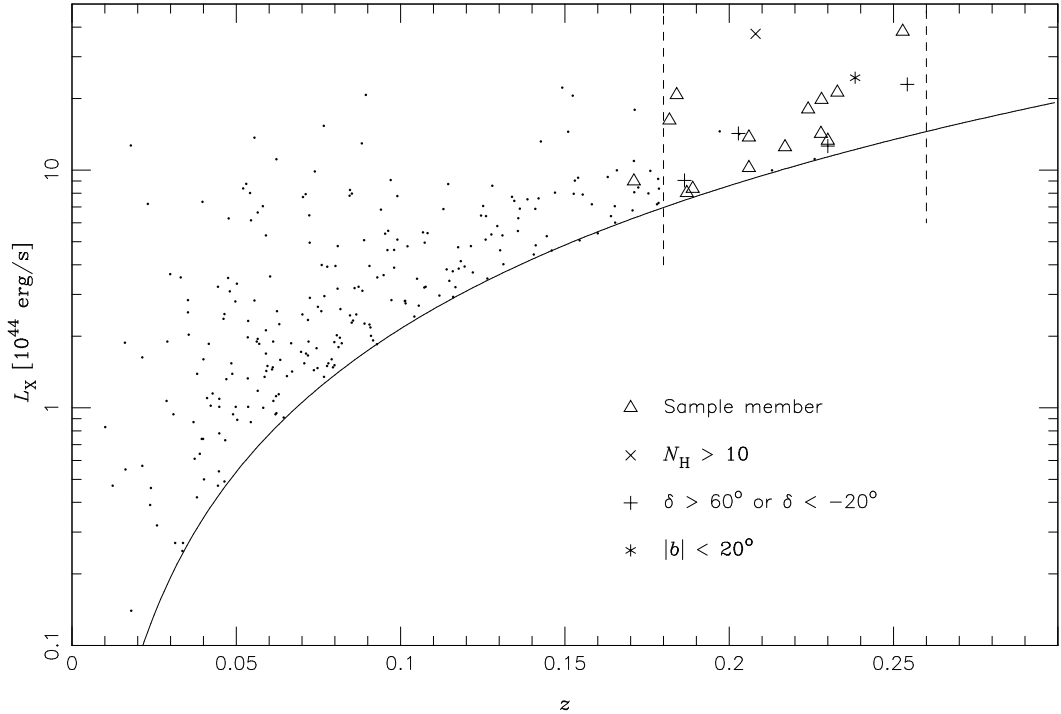}
  \end{center}
  \caption{Location of our sample within the XBACs
    catalogue. Cosmological parameters used are $h\!=\!0.5$, $\Omega_{\rm
      M}\! =\! 1$ and $\Omega_\Lambda\!=\! 0$. A~2218 falls outside the
      specified redshift range but will be observed in the same manner
      as the other clusters.}
  \label{fig:complete}
\end{figure}

Our sample is drawn from the XBACs catalogue~\cite{ebeling}, a
flux--limited catalogue of Abell clusters detected in the ROSAT
All--Sky Survey. We apply limits in redshift space of $0.18 < z <
0.26$ in order to have an approxiomately luminosity--limited
sample. It has been shown that X--ray luminosity is correlated with
cluster mass and therefore, given the difficulty of measuring X--ray
temperatures with current instruments, a luminosity--limited sample is
the best approximation to a mass--limited sample. The redshift range
has been chosen to maximize the lensing efficiency for a background
galaxy population at $\langle z\rangle = 0.8$~\cite{kneib}. Applying
further limits 
in declination (for accessibility from CFHT), galactic latitude (to
minimize contamination by stars) and hydrogen column density $N_{\rm
  H}$ leads to a sample of 14 clusters (listed in table~\ref{tab:sample}),
8 of which will be observed by HST under PI Kneib and 6 of which are
available in the HST archive. Figure \ref{fig:complete} shows that our
sample covers the corresponding region in the XBACs catalogue well. 

\begin{table}[ht]
  \caption{Physical data for our sample. Note that the temperatures
    given are {\it estimated}$\,^6$. The first eight clusters are observed
    with HST under PI Kneib, observations of the remaining six clusters
    are available in the HST archive.}
  \label{tab:sample}
  \hspace{0.4cm}
  \begin{center}
    \begin{tabular}{|l@{ }r|@{\hspace*{0.3cm}}r@{.}l@{\hspace*{0.3cm}}|@{\extracolsep{0.8cm}}r@{\extracolsep{0cm}}@{.}l|@{\extracolsep{0.5cm}}r@{\extracolsep{0cm}}@{.}l|}
      \hline
    \rule[-3mm]{0mm}{8mm}&   & \multicolumn{2}{c|}{\hspace{-0.2cm}$z$} & \multicolumn{2}{c|}{\hspace*{-0.8cm}$L_{\rm X} [10^{44}\,\mbox{erg/s}]$} & \multicolumn{2}{c|}{\hspace*{-0.5cm}$T_{\rm X}\,\mbox{[keV]}$} \\
      \hline \hline
     \rule[0mm]{0mm}{5mm}Abell & 68 & 0 & 1889 & 8 & 36 & 7 & 7 \\
      Abell &  209 & 0 & 2060 & 13 & 75 & 9 & 6 \\
      Abell &  267 & 0 & 2300 & 13 & 32 & 9 & 4 \\
      Abell &  383 & 0 & 1871 & 8 & 03 & 7 & 5 \\
      Abell &  773 & 0 & 2170 & 12 & 52 & 9 & 2 \\
      Abell &  963 & 0 & 2060 & 10 & 23 & 8 & 4 \\
      Abell & 1763 & 0 & 2279 & 14 & 23 & 9 & 7 \\
      \rule[-3mm]{0mm}{0mm}Abell & 1835 & 0 & 2528 & 38 & 34 & 15 & 1 \\
      \hline
      \rule[0mm]{0mm}{5mm}Abell & 1689 & 0 & 1840 & 20 & 74 & 10 & 8 \\
      Abell & 2218 & 0 & 1710 & 8 & 99 & 6 & 7 \\
      Abell & 2219 & 0 & 2281 & 19 & 80 & 11 & 2 \\
      Abell & 2261 & 0 & 2240 & 18 & 06 & 10 & 8 \\
      Abell & 2390 & 0 & 2329 & 21 & 25 & 11 & 6 \\
      \rule[-3mm]{0mm}{0mm}Abell &  665 & 0 & 1818 & 16 & 22 & 8 & 3 \\
      \hline
    \end{tabular}
  \end{center}
\end{table}

\section{Observations}
To date (June 2000) five clusters have been observed with HST, three
more are scheduled for observation before the end of 2000; six cluster
observations are available in the HST archive. The
observations are done with the WFPC2 through the F702W filter, three
orbits are allocated for each cluster. The excellent spatial
resolution of these images allows precise modelling of the mass
distribution in the cluster centres ($\la 100\,h^{-1}\,\mbox{kpc}$)
due to their 
strong lensing effects. For most of the clusters in our sample no
giant arcs were known previous to these observations. Therefore, they
will constitute a valuable sample for investigating the probablity of
formation of giant arcs which depends strongly on the cosmological
parameters, in particular $\Omega_\Lambda$~\cite{bartelmann}.

On larger scales, observations with the CFH12k camera on CFHT will be
used to determine mass profiles out to $1.5\dots2 h^{-1}$~Mpc using
the systematic distortion of the background galaxies due to weak
lensing by the cluster potential. CFH12k observations were finished
in June 2000. During three observing runs (7 nights) 11 clusters were
observed in three bands (B, R, I), typically reaching a limiting
magnitude of 25 in R. Photometry in three filters allows robust
discrimination between cluster members and background galaxies. 

Seven of our eight core sample clusters have been allocated XMM time
in category B under PI Kneib, with integration times between 20 and 30
ksec. All the other clusters will be observed by XMM under different
PIs. The observations will be done with the EPIC camera with a field
of view of 3.8~Mpc at $z\!=\!0.2$; images will
allow study of the morphology of the X--ray surface brightness and
detection of significant substructure, spectra will permit precise
measurement of X--ray temperatures and temperature profiles. 

These core observations will be supplemented by spectroscopy of giant
arcs and cluster galaxies, as well as miscellaneous observations in
different wave bands, notably in the near infrared in order to permit
determination of photometric redshifts in the central cluster regions. 

\section{Conclusions}
At the time of writing (June 2000), most of the optical observations have
been finished and are currently being reduced. A first
paper describing and modelling several arcs and multiple image systems
in Abell 383 is in preparation. Previously unknown arcs have already been
found in Abell 68, 383, 773, 963 and 1835. X--ray observations will begin
after the end of the XMM calibration phase and should be finished by
mid-- to end--2001. 
Further projects studying similar cluster samples at redshifts $\sim
0.1$ and $\sim 0.4$ are in preparation. 

\section*{Acknowledgements}

This work has been supported by the French--UK programme "Alliance" under
contract No.\ 00161XM.\\
OC acknowledges financial support by the European Commission under
contract No.\ ER--BFM--BI--CT97--2471.


\section*{References}
\begin{thebibliography}{99}
\bibitem{eke}Eke, V. R., Cole, S., Frenk, C. S. 1996, \MNRAS {\bf
    282}, 263
\bibitem{nfw}Navarro, J. F., Frenk, C. S., White, S. D. M. 1997, \ApJ
    {\bf 490}, 493
\bibitem{moore}Moore, B., Governato, F., Quinn, T., Stadel, J., Lake,
    G. 1998, \ApJ {\bf 499}, L5 
\bibitem{bower}Bower, R. G. 1998, \MNRAS {\bf 288}, 355
\bibitem{kneib}Natarajan, P., Kneib, J.--P. 1997, \MNRAS {\bf 287},
    833
\bibitem{ebeling}Ebeling, H., Voges, W., B\"ohringer, H., Edge, A. C.,
    Huchra, J. P., Briel, U. G. 1996, \MNRAS {\bf 281}, 799
\bibitem{bartelmann}Bartelmann, M., Huss, A., Colberg, J. M., Jenkins,
    A., Pearce, F. R. 1998, \AA {\bf 330}, 1
 
\end{thebibliography}
\end{document}